\documentclass[manuscript, screen]{acmart}

\usepackage{amsmath,amsfonts}
\usepackage{algorithmic}
\usepackage{graphicx}
\usepackage{textcomp}
\usepackage{xcolor}
\usepackage{xspace}
\usepackage{enumitem} 
\usepackage{xcolor}
\usepackage{listings, multicol}
\usepackage{threeparttable}
\usepackage[ruled,norelsize,linesnumbered]{algorithm2e}
\usepackage{longtable}
\usepackage{subcaption}
\usepackage[most]{tcolorbox}

\usepackage[normalem]{ulem}
\usepackage{tabularray}  
\usepackage{hyperref}

\newcommand{\sys}{\textsc{WizardMerge}\xspace}

\makeatletter
\newcommand{\removelatexerror}{\let\@latex@error\@gobble}
\makeatother

\newcommand{\lstbg}[3][0pt]{{\fboxsep#1\colorbox{#2}{\strut #3}}}
\definecolor{codegreen}{rgb}{0,0.6,0}
\lstdefinelanguage{diff}{
	frame=shadowbox,
	basicstyle=\ttfamily\scriptsize\bfseries,
        breaklines=true,
	morecomment=[f][\color{red}]{---}, 
	morecomment=[f][\color{codegreen}]{+++},
        morecomment=[f][\color{orange}]{<<<<<<<},
        morecomment=[f][\color{orange}]{=======},
        morecomment=[f][\color{orange}]{>>>>>>>},
	morecomment=[f][\lstbg{red!20}]{-\ },
	morecomment=[f][\lstbg{green!20}]{+\ },
	morecomment=[f][\color{blue}]{@@},
}

\newenvironment{mybullet}{\begin{list}{$\bullet$}
		{\setlength{\topsep}{0.5mm}\setlength{\itemsep}{0.5mm}
			\setlength{\parsep}{0.5mm}
			\setlength{\itemindent}{0.5mm}\setlength{\partopsep}{0.5mm}
			\setlength{\labelwidth}{15mm}
			\setlength{\leftmargin}{4mm}}}{\end{list}}

\usepackage[framemethod=TikZ]{mdframed}
\usepackage{lipsum}
\mdfsetup{skipabove=7pt,skipbelow=0pt}
\definecolor{darkcyan}{HTML}{0091A4}  
\definecolor{brightcyan}{HTML}{dcf0f2}
\newcounter{conclusioncounter}
\mdfdefinestyle{ccl}{
    linecolor=darkcyan, 
    outerlinewidth=3pt,
    leftline=true, 
    rightline=false, 
    topline=false, 
    bottomline=false, 
    innertopmargin=10pt, 
    innerbottommargin=10pt, 
    innerrightmargin=13pt, 
    innerleftmargin=10pt, 
    backgroundcolor=brightcyan 
}
\newcommand{\ccl}{\refstepcounter{conclusioncounter}\textbf{Conclusion:}\xspace}

\begin{document}

\title{WizardMerge - Save Us From Merging Without Any Clues}

\author{Qingyu Zhang}
\affiliation{%
  \institution{The University of Hong Kong}
  \city{Hong Kong}
  \country{China}}
\email{z1anqy@connect.hku.hk}

\author{Junzhe Li}
\affiliation{%
  \institution{The University of Hong Kong}
  \city{Hong Kong}
  \country{China}}
\email{jzzzli@connect.hku.hk}

\author{Jiayi Lin}
\affiliation{%
  \institution{The University of Hong Kong}
  \city{Hong Kong}
  \country{China}}
\email{linjy01@connect.hku.hk}

\author{Jie Ding}
\affiliation{%
  \institution{The University of Hong Kong}
  \city{Hong Kong}
  \country{China}}
\email{dingjie999888@icloud.com}

\author{Lanteng Lin}
\affiliation{%
  \institution{Beijing University of Posts and Telecommunications}
  \city{Beijing}
  \country{China}}
\email{lantern@bupt.edu.cn}

\author{Chenxiong Qian}
\authornote{Corresponding author.}
\affiliation{%
  \institution{The University of Hong Kong}
  \city{Hong Kong}
  \country{China}}
\email{cqian@cs.hku.hk}


\begin{abstract}

Modern software development necessitates efficient version-oriented collaboration among developers.
While Git is the most popular version control system, it generates unsatisfactory version merging results due to textual-based workflow, leading to potentially unexpected results in the merged version of the project.
Although numerous merging tools have been proposed for improving merge results, developers remain struggling to resolve the conflicts and fix incorrectly modified code without clues.
We present \sys, an auxiliary tool that leverages merging results from Git to retrieve code block dependency on text and LLVM-IR level and provide suggestions for developers to resolve errors introduced by textual merging.
Through the evaluation, we subjected \sys to testing on 227 conflicts within five large-scale projects. 
The outcomes demonstrate that \sys diminishes conflict merging time costs, achieving a 23.85\% reduction. 
Beyond addressing conflicts, \sys provides merging suggestions for over 70\% of the code blocks potentially affected by the conflicts. 
Notably, \sys exhibits the capability to identify conflict-unrelated code blocks that require manual intervention yet are harmfully applied by Git during the merging.
\end{abstract}




\maketitle

\section{Introduction}
\label{sec:intro}

As the most globally renowned version control system, Git~\cite{git} has been helping with the development and maintenance of millions of projects.
To automatically merge the codes from two different versions, Git utilizes a Three-way Merge Algorithm~\cite{three-way-merge} as the default strategy on the text level.
The textual-based merging is general and suitable for various types of text files and the merging tempo is proved to be the fastest.
However, it simply considers lines of code, neglecting all syntax and semantic information, which leads to potential bugs in the merged version of the project~\cite{compile-error-git-merge, runtime-error-git-merge}.

%
To mitigate the shortcomings of traditional textual-based merging, multiple merging approaches~\cite{automerge, mastery, fstmerge, spork, intellimerge, safemerge, jdime} are proposed for enhancing the correctness of merging results. 
These tools can be classified as \textbf{Structured Merging} and \textbf{Semi-structured Merging}.
Structured merging tools~\cite{spork, mastery, safemerge, automerge} transform the code intended for merging into an Abstract Syntax Tree (AST), thereby converting the code merging into the merging of edges and nodes on the AST~\cite{accioly2018understanding}. 
Semi-structured merging~\cite{jdime, fstmerge, intellimerge}, akin to structured merging, designates certain semantically insensitive AST nodes (e.g., strings) as text, allowing them to be directly processed by the text-based merging strategy during the merging process~\cite{cavalcanti2017evaluating}.


%
Despite structured and semi-structured merging tools facilitating code merging, challenges arise when it is unclear which version of the code should take precedence.
This uncertainty leads to merging conflicts. In the context of large-scale software projects, conflicts can become a source of frustration for developers, as they must dedicate significant effort to resolve the issues that arise from merging before they can release the final merged version~\cite{empiricalStudy, mergeWithChangeHistory}.
Addressing these conflict resolutions has given rise to merging assistance systems. 
These systems are designed with objectives such as selecting the most suitable developers to resolve conflicts~\cite{tipmerge}, prioritizing the resolution order for conflicts~\cite{somanyconflicts}, and even automatically generating alternative conflict resolutions~\cite{automerge}.
%
With the rapid advancements in machine learning technologies in recent years, researchers have turned to machine learning approaches~\cite{zhang2022using, MergeBERT, elias2023towards, dongmerge, aldndni2023automatic, deepmerge, mergegen}. 
These systems are trained using extensive datasets containing merge conflict scenarios and can provide predictions on the most suitable resolution strategy for each conflicting portion of code.

%
While these researches have made noteworthy contributions to code merging, they still exhibit the following limitations.
Firstly, structured~\cite{jdime, spork, mastery, safemerge, automerge} and semi-structured~\cite{fstmerge, intellimerge} still generate conflicts when the direct merging between two AST noes failed, and cannot assist developers in resolving them.
Secondly, conflict resolution assistance systems~\cite{somanyconflicts, automerge, tipmerge} focus solely on resolving conflicts, overlooking potential errors introduced by incorrectly applied but conflict-free code.
Though machine learning approaches~\cite{zhang2022using, MergeBERT, elias2023towards, dongmerge, aldndni2023automatic, deepmerge} can automate conflict merging, the functionality of the resulting code may not ensure alignment with developers' expectations. 
Additionally, machine learning methods necessitate specific code datasets for pre-training, which raises universality issues when applied to more diverse code projects.
Moreover, the developers often have unpredictable operations on resolving conflicts, where the auto-generated conflict resolution may fail to meet the expectations of them.

%
To address these limitations, we introduce a novel approach aiming to aid developers in handling complex code-merging scenarios.
Recognizing that \textbf{developers have unpredictable ultimate needs}, our approach focuses on grouping conflicts by relevance and providing prioritized suggestions for resolving conflicts rather than resolving them directly.
Our approach also detects the incorrectly applied codes as they can be easily neglected but cumbersome to handle for developers.
To realize these design goals, our approach requires a thorough understanding of the dependency relations in the two merging candidate versions.
Firstly, we perform definition dependency graph construction on LLVM Intermediate Representation (IR)~\cite{llvm} and utilize definition indicators extracted from Clang~\cite{llvm_clang} to indicate the definition name and ranges in the source file.
Following graph construction, our approach analyzes the Git merge outcomes that present all the conflicts and modified code blocks. 
Then, we align the definitions in the source file with code snippets in Git's outcomes to map from LLVM-IR to the Git merge results.
Given the dependency graph analysis and aligned Git merge results, we further conduct an edge-sensitive algorithm to detect incorrectly applied codes, which we deem as conflicts since they require developers' attention as well.
Finally, we assign priorities to these conflicted nodes and produce merging suggestions.

We implement a prototype \sys for C/C++ projects as most large-scale projects are written in C/C++~\cite{largest_projects}.
%
We evaluate \sys on 227 conflicts from five popular large projects. 
The outcomes demonstrate that \sys diminishes conflict merging time costs.
Beyond addressing conflicts, \sys provides merging suggestions for most of the code blocks affected by the conflicts. 
Moreover, \sys exhibits the capability to identify conflict-unrelated code blocks which should require manual intervention yet automatically applied by Git.

%
In summary, we make the following contributions:

\begin{mybullet}  

    \item We proposed a novel approach to infer the dependency between text-level code blocks with modification. 
    
    \item We proposed a remarkable methodology, which aims to pinpoint code blocks that have been automatically applied by Git but introduce unexpected results.
    
    \item We implemented a code merging auxiliary prototype named \sys based on the approaches and evaluated the system on projects with large-scale code bases. We will open source \sys and evaluation datasets at https://github.com/aa/bb.
\end{mybullet}

\section{Background}
\label{sec:bg}


\subsection{Git Merging}
\label{subsec:gitmerging}

\begin{figure}[th]
	\centerline{\includegraphics[width=\textwidth]{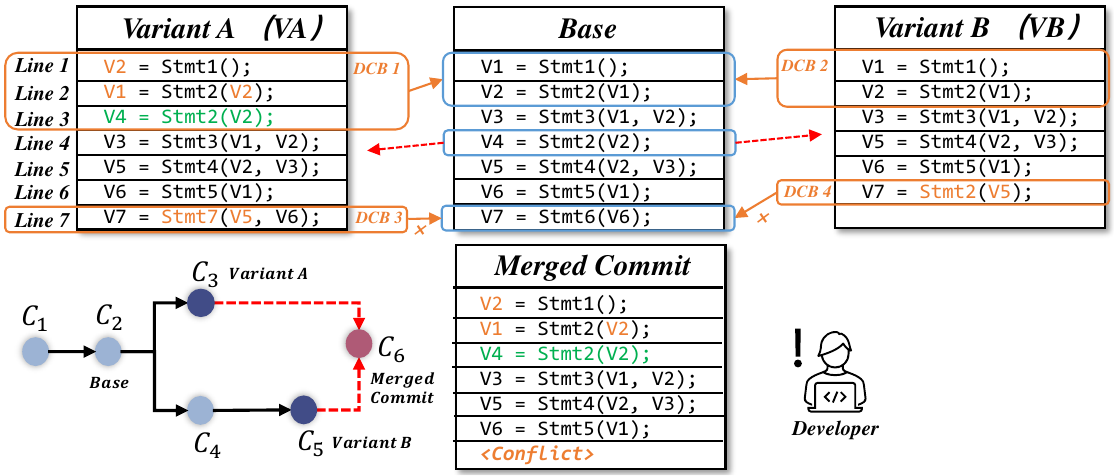}}
	\caption{\textmd{Three-way merging workflow of Git.}}
	\label{fg:three-way-merging}
\end{figure}

Git is a free and open-source distributed version control system designed for efficient handling of projects of all sizes~\cite{git}. 
Git merge, as a key operation, facilitates the integration of changes between versions and is often used to consolidate code changes, bug fixes, or new features~\cite{gitmergedocument}.
%
Git supports multiple textual-based merging algorithms~\cite{gitmergealgorithm}.
Take the default and most commonly employed algorithm, Three-way Merging Algorithm~\cite{three-way-merge} as an example, the workflow is shown in ~\autoref{fg:three-way-merging}.
When the developer tries to merge two commits (i.e., $C_3$ \textbf{Variant A (VA)} and $C_5$ \textbf{Vairant B (VB)} in ~\autoref{fg:three-way-merging}), Git first traverses the commit-graph back until it reaches the nearest common ancestor commit (i.e., $C_2$, named \textbf{Base}). 
Then, all source code blocks of each newer version are mapped to the corresponding code blocks in base versions, and the modified code blocks of either VA or VB are regarded as \textbf{Diff Code Blocks (DCBs)}.
A DCB in a variant may contain zero or multiple lines of code, indicating the removal of an existing line or modification/addition of multiple continuous lines.
After the code matching stage, Git will iterate each code block tuple, and determine which commit should be applied according to the following rules:

\begin{mybullet}
    \item If code blocks of \textit{VA} and \textit{VB} remain the same as the code block of \textit{Base}, or both of them are changed in the same way, inherit the code block from \textit{VA}.
    In~\autoref{fg:three-way-merging}, both VA and VB remove the definition of \textit{v4} (i.e., deleting line 4 of Base). 
    Due to identical modifications at the same locations, Git applies these changes. 
    It is noteworthy that even for code removal, VA and VB maintain a pair of empty DCBs, symbolizing the deletion of lines. 
    
    \item If only one code block of either \textit{VA} or \textit{VB} is changed, Git applies this code block, and marks this code block and the corresponding code block in the other variant as DCBs. 
    In~\autoref{fg:three-way-merging}, VA alters lines 1-2 and introduces a new line at 3. 
    Git's code-matching algorithm identifies these modifications, designating lines 1-3 of VA as \textit{DCB 1} and lines 1-2 of VB as \textit{DCB 2}. 
    Since only \textit{DCB 1} changes, Git applies it to generate the code for the Merged Commit at lines 1-3. 

    \item If both code blocks from \textit{VA} and \textit{VB} are changed, and the modifications are different, Git marks both code blocks as DCBs.
    As Git cannot decide which one should be applied, it emits a conflict containing the two DCBs.
    In~\autoref{fg:three-way-merging}, VA and VB independently modify the definition of \textit{v7}. Git detects these alterations, marking line 7 of VA as \textit{DCB 3} and line 6 of VB as \textit{DCB 4}.
    Since these two modifications differ, Git cannot determine which one to apply. 
    Consequently, Git signals the inconsistency between \textit{DCB 3} and \textit{DCB 4}, prompting developers to resolve this conflict.
    
\end{mybullet}

\begin{figure}[th]
        \vspace{-0.5\baselineskip}
	\centerline{\includegraphics[width=0.9\linewidth]{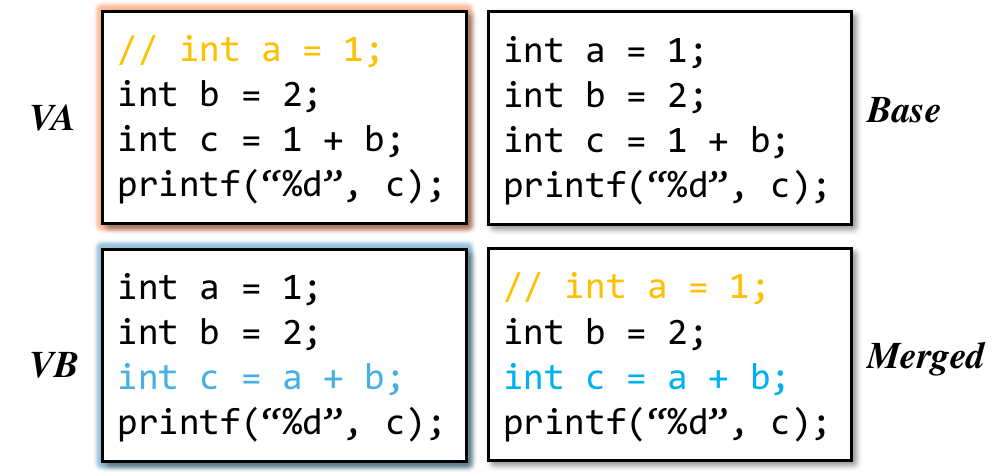}}
        \vspace{-0.3\baselineskip}
	\caption{\textmd{An example of a clean but incorrect Git merge with the three-way merging algorithm}}
	\label{fg:vDCB_example}
        \vspace{-0.7\baselineskip}
\end{figure}

\subsection{Limitations of Existing Approaches}
\label{subsec:limitation}

Despite being the most widely used code merging tool currently, Git merge faces the following two challenges: 
\textit{(Challenge 1) It cannot ascertain developers' expectations for code functionality.} Consequently, when both versions modify the same code, Git treats this discrepancy as a conflict, necessitating manual intervention for resolution. We discussed how the conflicts are generated in \S \ref{subsec:gitmerging}.
\textit{(Challenge 2) It cannot ensure the correctness of successfully applied and conflict-free code blocks.} Even when Git successfully merges two versions without conflicts, the merged code may be generated with new bugs. 
~\autoref{fg:vDCB_example} shows an incorrect merging scenario.
Only the first line of \textit{VA} and the third line of \textit{VB} are changed.
Using Git merge, the two versions of code can be easily merged without any conflict (i.e., the first line applied from VA and the second line applied from VB).
Unfortunately, in the merged version, the definition of variable \textit{a} has been removed, which causes a compilation error at the third line.  
These DCBs enforced by Git without conflict, yet resulting in unexpected results, are termed \textbf{Violated DCBs}.
Given that these violated DCBs are not flagged as conflicts and are automatically incorporated by Git merge, they tend to introduce concealed programming errors into the merged code. 
In the most severe scenarios, these error-prone code segments might seemingly execute successfully but could potentially give rise to security issues post-deployment~\cite{compile-error-git-merge, runtime-error-git-merge}.

\begin{figure}[th]
\centerline{\includegraphics[width=\linewidth]{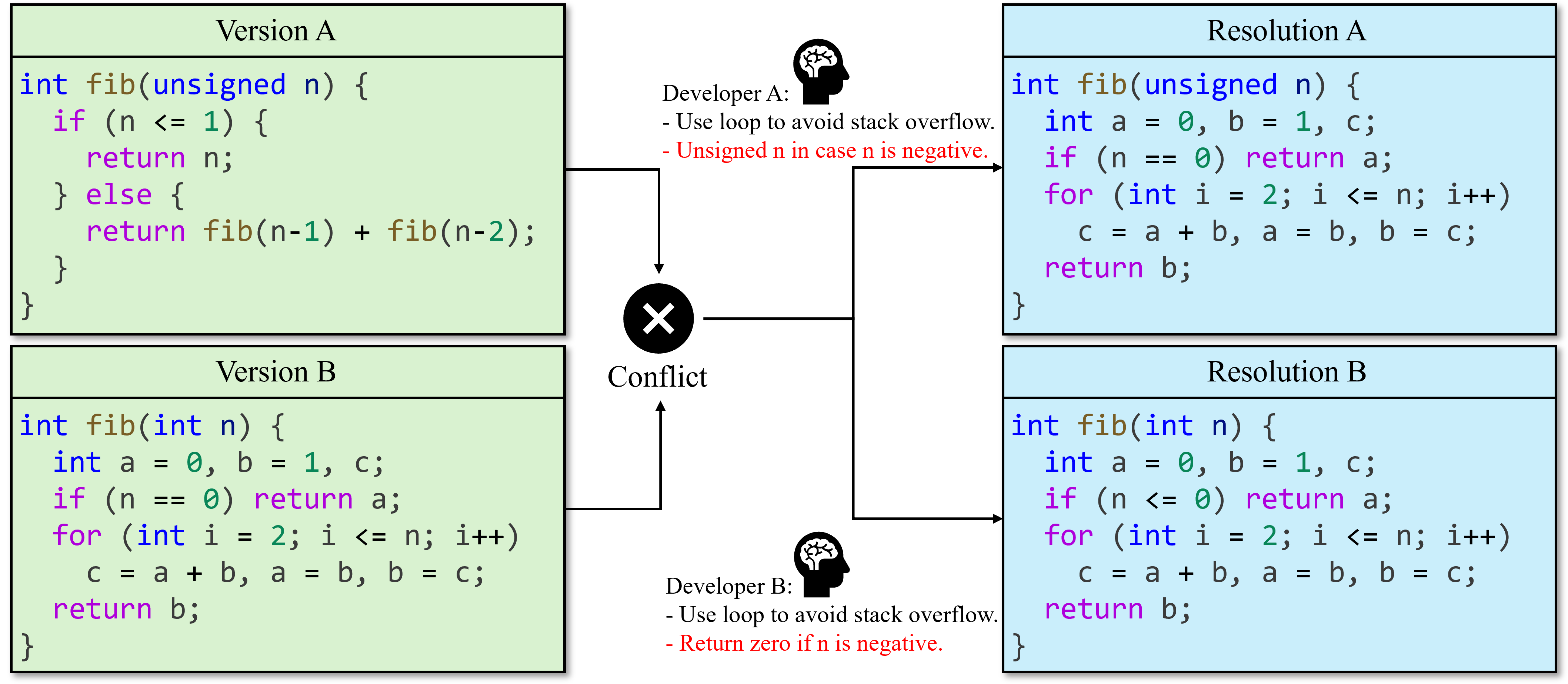}}
	\caption{\textmd{An example of different resolutions to the same conflict}}
	\label{fg:diff_resolution}
\end{figure}

Over the years, researchers have concentrated on enhancing the quality of merged code and conflict resolution. 
However, these two basic limitations have not yet been fully resolved.
On the one hand, structured~\cite{mastery, spork, safemerge, automerge} and semi-structured~\cite{jdime, fstmerge, intellimerge} merging still report conflicts when the same AST node is modified simultaneously by two merging candidate branches.
On the other hand, the conflict resolution assistance system~\cite{somanyconflicts, automerge} and machine learning approaches~\cite{zhang2022using, MergeBERT, elias2023towards, dongmerge, aldndni2023automatic, deepmerge, mergegen} only consider conflicts reported by the merge tool, ignoring all code snippets that have been applied.
Additionally, automatic software merging tools~\cite{mastery, spork, safemerge, automerge, jdime, fstmerge, intellimerge, zhang2022using, MergeBERT, elias2023towards, dongmerge, aldndni2023automatic, deepmerge, mergegen} neglect an essential observation: \textbf{developers have unpredictable ultimate needs}, therefore cannot fully overcome \textit{Challenge 1}.
To better illustrate this, we provide an example of implementing the Fibonacci sequence in \autoref{fg:diff_resolution}. 
Two versions, A and B, calculate the Fibonacci sequence using recursion and iteration, respectively.
When attempting to merge these versions, a conflict arises, and two developers resolve it. Although both developers decide to use the iterative approach from version B to avoid stack overflow for large values of $n$, they implement different strategies for handling negative values of $n$.
Developer A, inspired by version A, uses an unsigned definition for 
$n$ to eliminate negative values (line 1 in resolution A). In contrast, Developer B opts to return zero when $n$ is less than zero (line 3 in resolution B).
Since developers' decisions on conflict resolution are unpredictable, automatic software merging tools often fail to produce an acceptable resolution in such cases.

Furthermore, machine learning systems~\cite{zhang2022using, MergeBERT, elias2023towards, dongmerge, aldndni2023automatic, deepmerge, mergegen} are limited in their ability to resolve conflicts in large codebases due to input and output sequence constraints. 
However, these conflicts typically do not significantly impact developers' time. 
For example, even MergeGen~\cite{mergegen}, the most advanced system available, failed to achieve satisfactory results when applied to the large-scale targets evaluated in \S\ref{sec:eval}. 
Upon completing MergeGen's pre-training, we discovered that it produced no correct outputs, achieving an accuracy of 0\%. 
This failure can be attributed to MergeGen's default input and output sequence length limitations of 500 and 200 BPE, respectively~\cite{mergegen}. 
These limitations are excessively stringent for large-scale project source code, leading to truncation of code beyond the prescribed lengths. 
However, the end of the truncated code is usually far away from where conflict occurs and MergeGen cannot master the ability to resolve the conflicts given the trimmed training and testing sets.
Attempts to relax these length limitations are impractical due to increased GPU memory consumption during training. 
Moreover, while machine learning techniques excel in generating merge results, they may not always align perfectly with developers' expectations. 
Although MergeGen's evaluation results demonstrate superior conflict code matching precision and accuracy on a line of code basis compared to previous works~\cite{mergegen}, it still falls short of expectations in some cases, with average precision and accuracy rates of 71.83\% and 69.93\%, respectively.
\section{Design}
\label{sec:design}
Given the challenges outlined in the preceding section, we have identified two key insights. 
Addressing \textit{Challenge 1}, we recognize that the expected outcomes of merges are subjective and open-ended. 
Consequently, our approach emphasizes guiding developers in resolving conflicts rather than prescribing resolutions. 
To tackle \textit{Challenge 2}, it is imperative to analyze all modified code.
To this end, we present \sys, a novel conflict resolution assistance system, which provides developers with clues and aids in conflict resolution. 
Moreover, \sys emphasizes all modified code snippets, not just conflicts, enabling it to uncover hidden discrepancies and preempt potential errors. 
A detailed design of \sys will be presented in the subsequent subsections.

\subsection{Overview}

\begin{figure}[h]
	\centerline{\includegraphics[width=\linewidth]{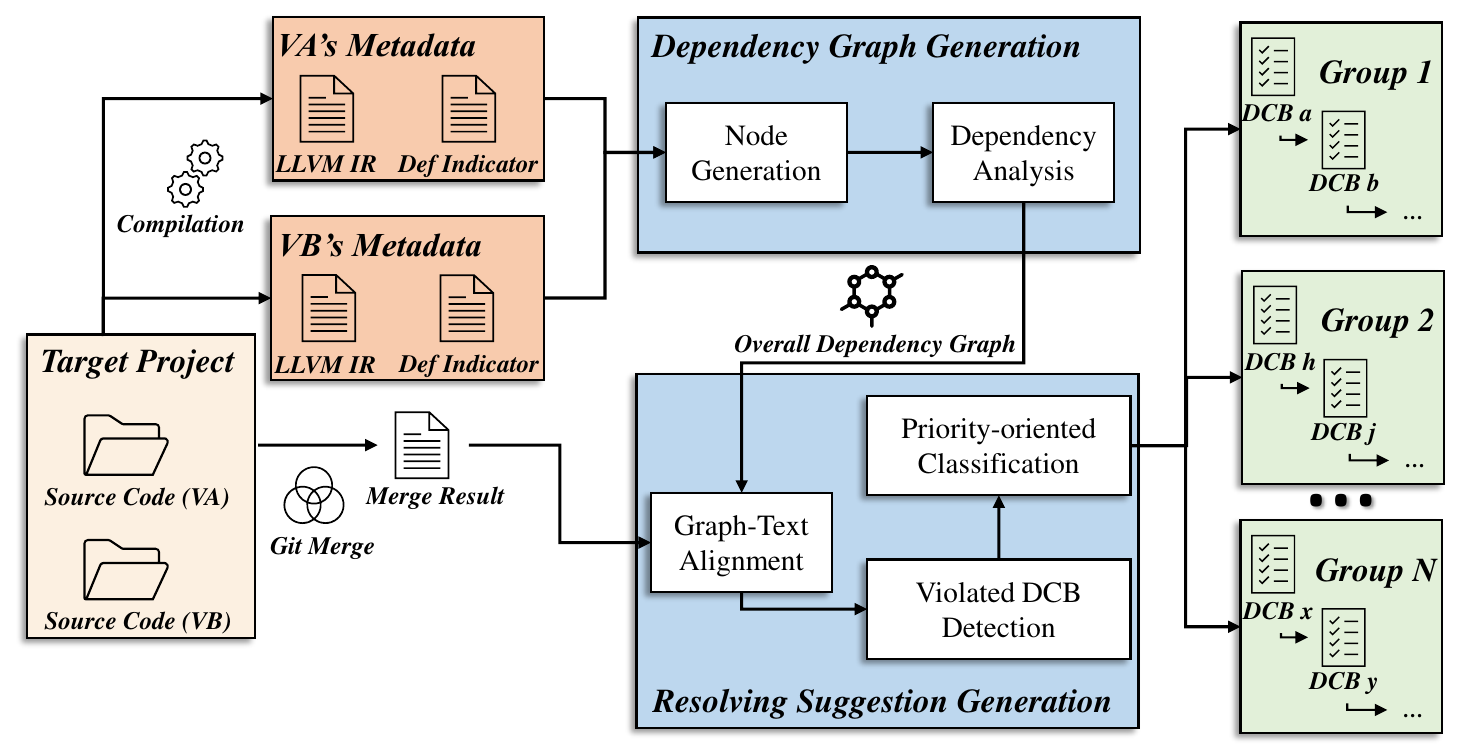}}
	\caption{\textmd{System overview of \sys.}}
	\label{fg:overview}
        \vspace{-\baselineskip}
\end{figure}

\autoref{fg:overview} depicts the system overview of \sys. 
\sys first consumes compilation metadata from the two merge candidates through LLVM\cite{llvm}.
\sys employs LLVM based on two insights: 
\romannumeral1) For large projects, code is compiled before a new version release to ensure compilation success, allowing \sys to collect metadata without introducing additional merging steps;
\romannumeral2) LLVM Intermediate Representation (IR) provides a robust static analysis framework and comprehensive debug information, facilitating \sys in performing definition dependency reasoning.
To bridge LLVM IR analysis with merge results from Git, the compilation metadata also encompasses definition indicators.
A definition indicator serves as the representation of a named definition section at the source code level, indicating the range and name of a definition. 
For example, a definition indicator of a function records its function name, along with its start and end line numbers in the source code file. Subsequently, 
To identify relationships between all modified codes, \sys advances to the \textbf{Dependency Graph Generation} stage. 
Instead of focusing only on conflicts, dependencies for all definitions are analyzed.
\sys creates vertices for the definitions and establishes edges between vertices according to the dependency relationships of definitions analyzed from LLVM IRs. 
Following this, the \textbf{Overall Dependency Graph (ODG)} is generated to represent the dependencies among all the definitions in both candidate versions of the target project.
To attach Git's textual results with IR-level dependency information, \sys aligns Git merge's results and ODG to match DCBs with vertices in ODG through \textbf{Graph-Text Alignment}.
This stage is capable of mapping DCBs to the corresponding definition nodes in each ODG, refining definition nodes into DCB nodes, and transforming definition dependencies into DCB dependencies.
To rectify errors introduced by Git merge, \sys employs an edge-sensitive algorithm in the \textbf{Violated DCB Detection} stage, detecting both conflict-related and conflict-unrelated violated DCBs. 
These violated DCBs and conflicts are categorized, prioritized based on the dependency graph, and presented as suggestions to developers to assist in resolution.
\vspace{-0.5\baselineskip}

\subsection{Dependency Graph Generation}
\label{subsec:dgg}
\sys first constructs ODG based on the LLVM IR and definition indicator.
\sys currently supports the following types of nodes and edges:

\textit{Graph node.}
Subsequently, \sys generates nodes for each named definition with the corresponding indicator.
Considering the types of definitions, \sys creates three types of nodes based on the indicator information: 
\textbf{Type Node (TN)} represents composite types, e.g. structure, class, type alias, and enumeration; 
\textbf{Global Variable Node (GN)} represents global variable;
\textbf{Function Node (FN)} represents function.

\textit{Graph edge.}
Each named definition may depend on other definitions within the project.
For instance, a function might utilize a global variable or return a pointer with a specific structure type. 
In the ODG, these usage dependencies among definitions are deduced from LLVM IRs and represented by the edges linking one node to another.
The types of edges are defined by the types of their two connected nodes.
An edge of type A-B represents a dependency from node A to node B.
In \sys, five classifications of edges are supported: 
TN-TN, GN-TN, FN-TN, FN-GN and FN-FN.

Note that \sys will never generate an edge from TN to FN, even though member functions may be defined within a composite type. 
Instead, \sys creates TN for all functions, irrespective of whether the function is standalone or nested within another scope. 
If an FN represents a nested function defined inside a composite type, an FN-TN edge is built to represent the dependency from the function to the specific type.
\vspace{-0.5\baselineskip}

\subsection{Resolving Suggestion Generation}
\label{subsec:rsg}
In this stage, \sys extracts DCB information from Git merge to aid in the analysis.
Then, merging suggestions can be generated based on the analysis results.

\subsubsection{Graph-Text Alignment}

The merge results cannot be directly linked to the corresponding node in ODG. 
This is primarily because these results are raw texts that provide no additional clues for program analysis. 
In the raw results, a single large DCB may cover multiple nodes and a single node may encompass several DCBs. 
This N-to-N relationship complicates the analysis of the graph.

\begin{figure}[th]
	\centerline{\includegraphics[width=0.93\linewidth]{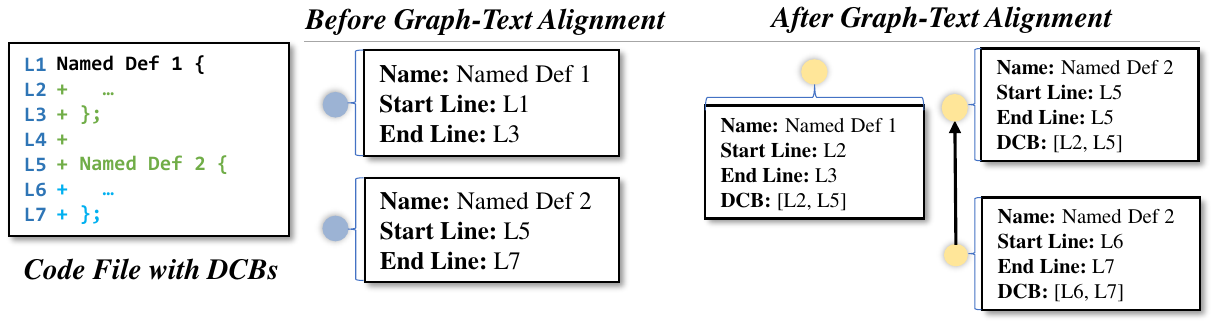}}
	\caption{\textmd{Example alignment between node and DCB text.}}
	\label{fg:gtalign}
\end{figure}

\sys initially associates each DCB with its corresponding definition indicators.
To identify all nodes contained in the file where the DCB is situated, \sys extracts the nodes whose range intersects with the DCB range. 
If no matched node is found, it indicates that the DCB is not in a definition (e.g., newlines or comments) and will not be further analyzed by \sys.
Treating each DCB as a "line segment", its start and end line numbers represent the coordinates of the line segment's two endpoints on the coordinate axis. 
Similarly, each definition indicator of a node is considered as a unique "color".
The range of each node signifies a specific coordinate interval that is colored with the definition represented by the node. 
Thus, retrieving the node during each DCB traversal can be seen as a color query for the interval $[start\_line, end\_line]$, and
interval modification and information maintenance can be efficiently managed using line segment trees.
For a target project containing $N$ lines of code, each "coloring" and "color query" operation can be achieved with a time complexity of $O(log_2N)$~\cite{segmenttree}.
\sys subsequently logs all DCBs corresponding to each node, organizing them in ascending order of line numbers. 
The node is then subdivided into finer-grained nodes. 
The ranges of these finer-grained nodes are updated concurrently based on the original node and their corresponding unique DCBs.
Fine-grained nodes belonging to the same definition will be connected sequentially in range order, which implicitly expresses the dependency between code blocks.
This mechanism ensures that each node in ODG aligns with at most one DCB, effectively preventing any matching confusion within the ODG.

Taking ~\autoref{fg:gtalign} as an instance, there are two definitions: \textit{Named Def 1}, \textit{Named Def 2}. 
Before alignment, the nodes are created with their range information. 
Let's consider two DCBs associated with these definitions ($[L2, L5]$ and $[L6, L7]$). 
To establish a one-to-one mapping between DCBs and nodes, each node can be divided into multiple ones, while unaffected lines are excluded as they do not influence the merging. 
For DCB $[L2, L5]$, \sys detects that $[L2, L3]$ belongs to \textit{Named Def 1} and $[L5, L5]$ belongs to \textit{Named Def 2}.
Consequently, a new node starting from $L2$ to $L3$ will replace the original node for \textit{Named Def 1}.
As $[L1, L1]$ remains unchanged, it will be discarded. 
Furthermore, a new node from $L5$ to $L5$ is also formed since $[L6, L7]$, representing another portion of \textit{Named Def 2}, corresponds to the DCB $[L6, L7]$. 
To indicate the code dependency from lines to previous lines within the same definition, an additional edge from node $[L6, L7]$ to node $[L5, L5]$ is established.

As \sys is tasked with managing two code versions (VA and VB), some nodes in the ODG of VA must possess a corresponding mirror node in the ODG of VB. 
\sys also constructs a mapping of each node to its matched node in the other version's ODG by matching the name of the definition and the DCB metadata linked to the node. 
Note that if one node is aligned with a DCB, it must have one unique mirror node with the other DCB in the same DCB pair.
If the node has a matched node, the matched node must be the mirror node and the definition name is consistent with the current node.
The matched node results contribute to the violated DCB detection via dependency missing determination.
Besides, the ODG's size can also be shrunk if there are nodes not attached to any DCBs, which means they are not changed during merging.
After the node deletion, the newly generated graph is called \textbf{Shrunk Dependency Graph (SDG)}.

\subsubsection{Violated DCB Detection}
\label{ssc:vioaltedDCBDetection}
Violated DCBs are the underlying cause of seemingly successful but erroneous merging.
To offer developers a comprehensive understanding of the merging process, \sys detects violated DCBs based on an edge-sensitive algorithm and deems them as conflicts requiring resolution suggestions. 
Given the SDGs and node matching relationship, \sys traverses each edge in the two versions of the SDG separately. 
Each edge, starting from $v$ and ending at $u$, signifies that $v$ relies on $u$.
\sys determines which version of the DCB associated with $u$ and $v$ is ultimately applied in the merge results (i.e., from VA, VB, or conflict).
\sys then ascertains whether these nodes represent violated DCBs by the applied versions and matching nodes of $u$ and $v$. 
We introduce all the safe (i.e. not violated DCBs detected) and violated scenarios as follows.

\paragraph{Symbols Definition}
Let the SDGs from VA and VB be represented as $G_A=\ <V(G_A), E(G_A)>$ and $G_B=\ <V(G_B), E(G_B)>$, where $V$ indicates the vertices set and $E$ indicates the edges set.
For each vertex $v \in V$, we represent its mirror node as $Mi(v)$ and use $match(Mi(v))$ to indicate whether the mirror node of $v$ is a matching node ($True$ or $False$).
$v$ belongs to only one of the $V_N$, $V_A$, and $V_C$ vertex sets, representing not applied, applied, and conflict node sets respectively.
For each edge $e(v, u) \in E$, it represents an edge from $v$ to $u$, indicating $v$'s dependency on $u$.
According to the construction principles of SDG, there are two basic facts:
\begin{mybullet}
    \item If $v \in V_A$, then $Mi(v) \in V_N$, and vice versa.

    \item If $v \in V_C$, then $Mi(v) \in V_C$.
\end{mybullet}
As each $v \in V_A$ must indicate another vertex (i.e., $Mi(v)$) belongs to $V_N$, \sys only analyzes the cases where $v \in V_A$ or $v \in V_C$ to avoid duplicate traversal.

\begin{figure}[th]
	\centerline{\includegraphics[width=0.99\linewidth]{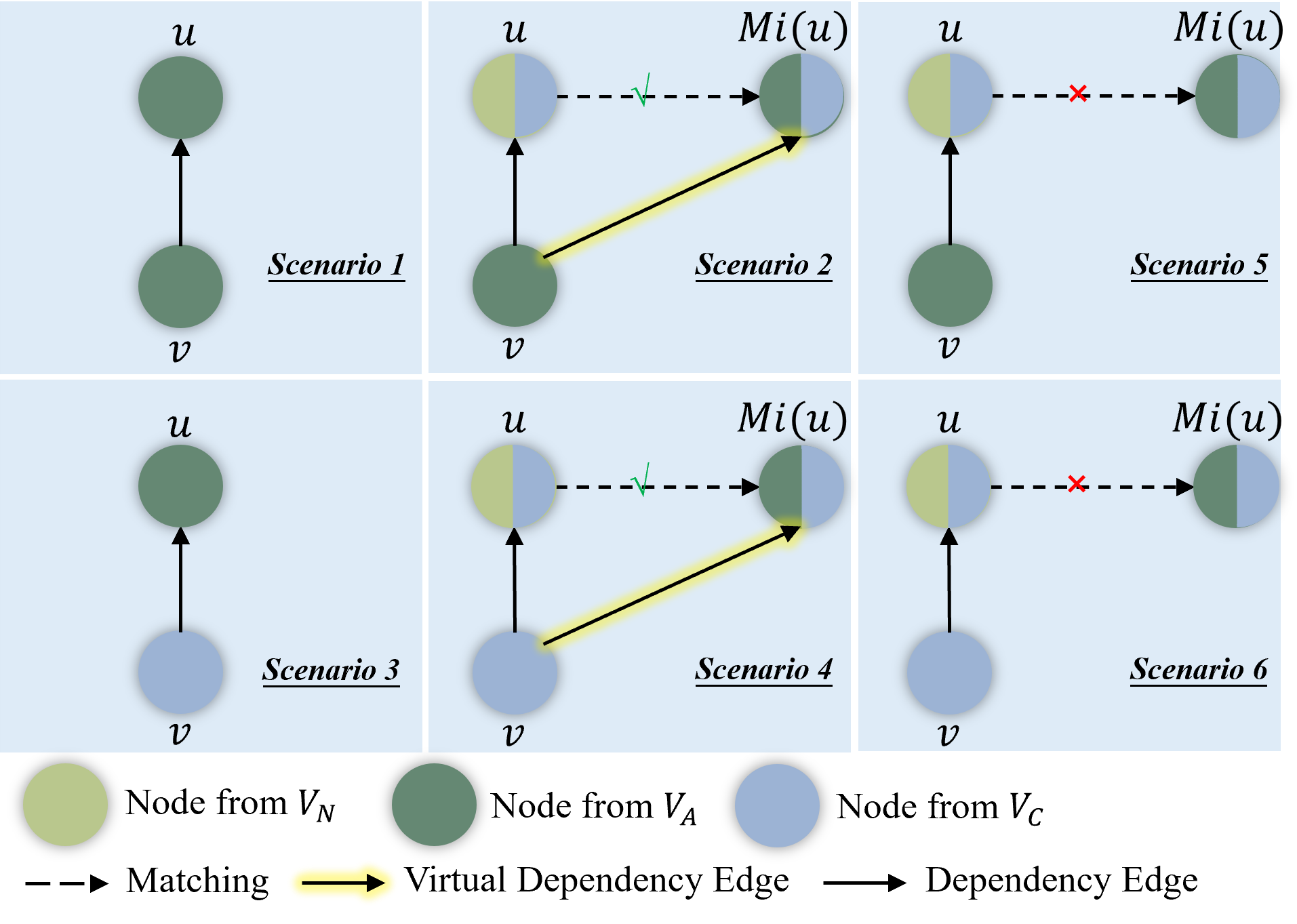}}
	\caption{\textmd{Edge traversal scenarios visualization}}
	\label{fg:all_sc}
\end{figure}

\paragraph{Safe Scenarios}
For each $v$ and an edge $e = (v, u)$, we define the scenarios satisfying any of the following four formulas as safe:
\begin{equation}\label{eq:safe_1}
v \in V_A, u \in V_A
\end{equation}
\begin{equation}\label{eq:safe_2}
v \in V_A, u \notin V_A, match(Mi(u)) = True
\end{equation}
\begin{equation}\label{eq:safe_3}
v \in V_C, u \in V_A
\end{equation}
\begin{equation}\label{eq:safe_4}
v \in V_C, u \notin V_A, match(Mi(u)) = True
\end{equation}
In SDGs, the above formulas can be regarded as an edge-sensitive detection algorithm, which is visiualized as depicted in ~\autoref{fg:all_sc}.
For \autoref{eq:safe_1}, if both $u$ and $v$ are applied, it means the dependency relation between nodes within the same SDG (as shown in \textit{Scenario 1} of ~\autoref{fg:all_sc}).
By the same token, \textit{Scenario 3} also illustrates such a direct dependency represented by ~\autoref{eq:safe_3}.
For \autoref{eq:safe_2} (\textit{Scenario 2} in ~\autoref{fg:all_sc}), $v$ is applied but $u$ is either not applied ($u \in V_N$) or in conflict status ($u \in V_C$).
Nevertheless, if the mirror of $u$ matches $u$ ($match(Mi(u)) = True$), it means that at least in the other SDG, $v$ can find its corresponding dependency applied (i.e., $Mi(u)$) although $u$ is not applied, or both versions of conflicted $u$ hold the same dependency for $v$.
To this end, \sys uses a virtual dependency edge from $v$ to $Mi(u)$ to indicate a cross dependency between nodes from different SDGs. 
Similarly, ~\autoref{eq:safe_4} (\textit{Scenario 4}) is also regarded as safe with the same reason.

\paragraph{Violated Scenarios}
For each $v$ and an edge $e = (v, u)$, all the scenarios not detected as safe are regarded as violated scenarios.
These scenarios are:
\begin{equation}\label{eq:violated_1}
v \in V_A, u \notin V_A, match(Mi(u)) = False
\end{equation}
\begin{equation}\label{eq:violated_2}
v \in V_C, u \notin V_A, match(Mi(u)) = False
\end{equation}
The \textit{Scenario 5} and \textit{Scenario 6} in ~\autoref{fg:all_sc} visualize ~\autoref{eq:violated_1} and ~\autoref{eq:violated_2} respectively.
Once $u$ is not applied or in conflict, and the mirror node of $u$ does not match it ($match(Mi(u)) = False$), it means that $v$ misses dependency from $u$ in both SDGs (when $u \in V_N$), or $Mi(u)$ holds a different dependency from $u$, which does not match $v$'s requirement (when $u \in V_C$).
\sys identifies these cases are incorrectly handled by Git merge and marks the corresponding $v$, $u$, $Mi(v)$ and $Mi(u)$ as violated.
Note that vertices with conflicts are always treated as violated, regardless of whether they are encountered in safe or violated scenarios.
Finally, the DCBs attached to these vertices are also marked as violated.

\begin{figure}[th]
        \vspace{-0.5\baselineskip}
	\centerline{\includegraphics[width=0.85\linewidth]{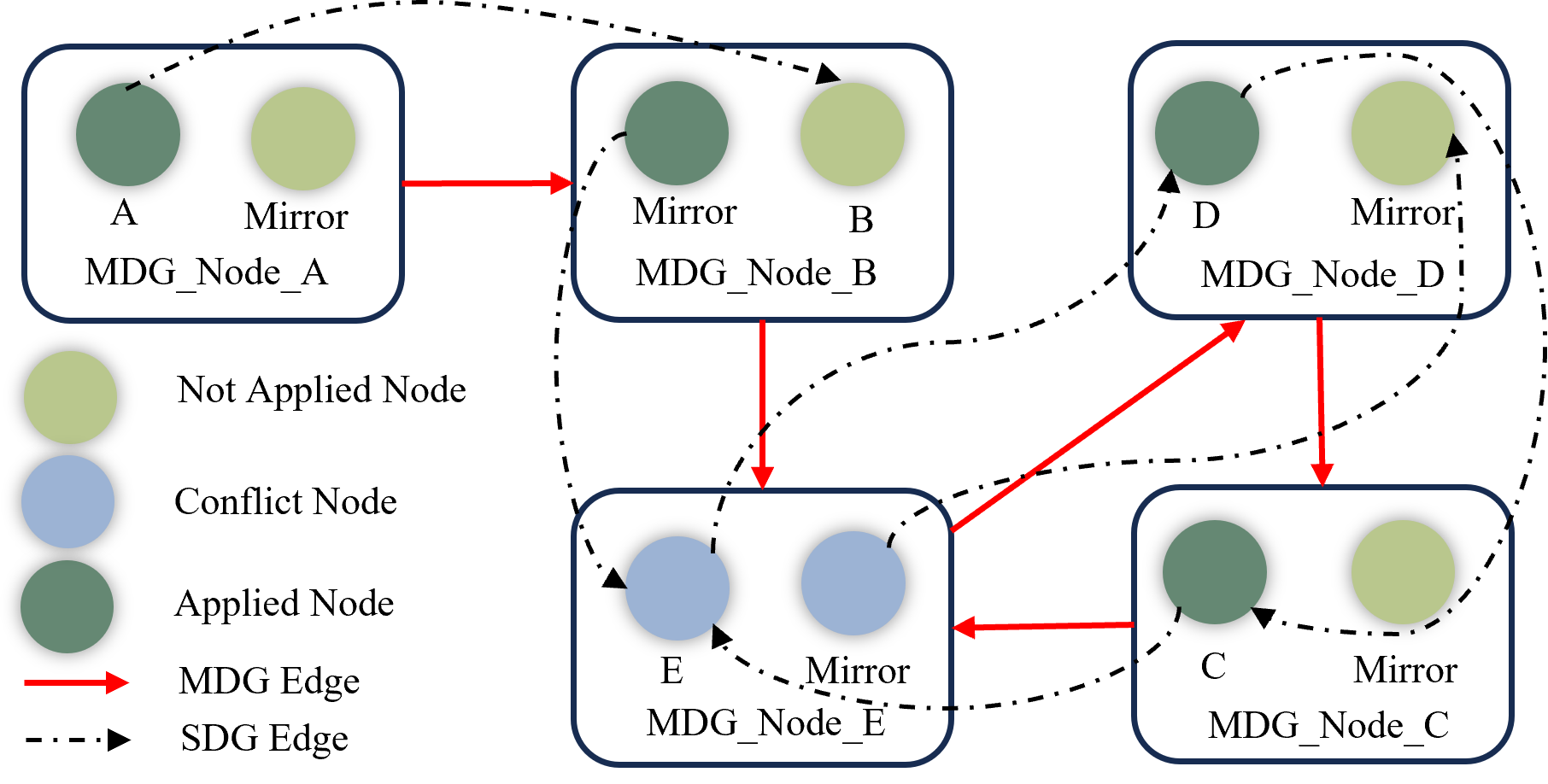}}
	\caption{\textmd{MDG example. Each node in MDG contains exactly a pair of mirror nodes in SDG. Each edge is built if there is at least one SDG edge from one SDG node to another.}}
	\label{fg:mdg}
\end{figure}

\subsubsection{Priority-Oriented Classification}
\sys further minimizes the number of nodes that require processing and constructs a \textbf{Minimum Dependency Graph (MDG)} to emulate the dependencies of the merged version of code. 
In the MDG, each node will consist of either an applied node or a conflict node and its matching node, and the edges between nodes will be created based on various SDGs according to the applying status. 
Specifically, while establishing the MDG, \sys will traverse each conflict node or node associated with violated DCBs, and based on the applying status of the node, it will switch to the corresponding version of SDG and continue traversing the nodes in that particular SDG. 
Particularly, if the node is in conflict status, then the outcoming edges of both the mirror of the node and itself would be iterated. 
Concurrently, \sys sets up a directed edge between corresponding nodes for the MDG, relying on the encountered edges during traversal. 
Ultimately, the MDG encapsulates the essential dependencies of the two versions of code to be merged.

\autoref{fg:mdg} shows an instance illustrating how MDG is built from SDG.
Each MDG node is created from a pair of nodes related to conflict or violated DCBs.
SDG node \textit{A-E} are from VA and all \textit{Mirror} nodes are from VB.
Starting from A and its mirror, \sys iterates the outcoming edges and finds B is linked to a violated DCB.
Then, \sys creates a new node \textit{MDG\_Node\_B} containing B and its mirror.
As B is not applied while its mirror node is applied, \sys iterates the outcoming edges of the mirror of B and finds a pair of conflict nodes (i.e., E and Mirror), which forms \textit{MDG\_Node\_E}. 
\sys then iterates the outcoming edges of both B and Mirror because of the conflict.
By the same token, \textit{MDG\_Node\_D} and \textit{MDG\_Node\_C} are created.
\sys then employs the Tarjan algorithm~\cite{tarjan} to condense the MDG nodes and eliminate cycles. 
During this process, nodes in the same strongly connected component are regarded as having identical priority, as these nodes may reference each other within the code. 
In the case of the directed acyclic MDG, \sys divides the MDG into multiple subgraphs based on connectivity. 
All nodes within each subgraph are grouped, and each node is assigned a priority through the topological sorting of the subgraph. 
Nodes that appear later in the topological order signify that other nodes have greater dependencies on them, while the nodes themselves have fewer dependencies. 
Consequently, these nodes are assigned higher priority, enabling developers to address them earlier.
In ~\autoref{fg:mdg}, \textit{MDG\_Node\_C}, \textit{MDG\_Node\_D} and \textit{MDG\_Node\_E} are in the same cyclic of MDG and are thus assigned the same resolution priority as they depend on each other.
After calculating the topological order, \sys suggests that the DCB resolution priority should be \{\textit{MDG\_Node\_C, MDG\_Node\_D, MDG\_Node\_E}\}, \{\textit{MDG\_Node\_B}\} and \{\textit{MDG\_Node\_A}\}.

\section{Evaluation}
\label{sec:eval}
In this section, we evaluate the effectiveness of \sys, aiming to answer the following research questions:

\begin{mybullet}
    \item \textbf{RQ1: } Can \sys help reduce the time consumption of merging conflicts?

    \item \textbf{RQ2: } Can \sys detect violated DCBs related to conflicts applied by Git? 

    \item \textbf{RQ3: } Can \sys detect violated DCBs unrelated to conflicts applied by Git? 

\end{mybullet}

\textit{Dataset Collection.}\label{pg:dataset}
%
We choose Firefox~\cite{firefox}, Linux Kernel~\cite{linux-kernel}, MySQL~\cite{mysql}, PHP~\cite{mysql}, and LLVM~\cite{llvm} as the target projects because of their widespread usage as open-source software. 
Additionally, they boast extensive code bases and are maintained by multiple developers, resulting in repositories that feature intricate commits and merging scenarios.
We collect 22 conflict pairs from Firefox, 43 from Linux Kernel, 17 from LLVM, 46 from PHP, and 99 from MySQL with a committing timeframe from 2021 to 2024.
For each pair of conflicts, we only focus on the files in C/C++ format and guarantee they either have at least five conflict regions or violated DCBs detected.
On average, a conflict contains 824.33 lines of codes and is related to 6.8 files.

%
\textit{Evaluate with State-of-the-art (SOTA) Systems.}
We initially intended to assess \sys alongside other cutting-edge systems.
However, JDIME~\cite{jdime}, Mastery~\cite{mastery}, Spork~\cite{spork}, SafeMerge~\cite{safemerge}, AutoMerge~\cite{automerge}, IntelliMerge~\cite{intellimerge} and SoManyConflicts~\cite{somanyconflicts} encountered language incompatibility issues within our datasets.
Besides, migrating \sys to the languages these systems support is impractical because as an LLVM-based project, \sys requires the full support from the LLVM framework, while the compiler front ends of Java~\cite{llvm_java_front}, Typescript~\cite{llvm_ts_js_front}, and Javascript~\cite{llvm_ts_js_front} are still under development. 
Furthermore, FSTMerge\cite{fstmerge} purports scalability contingent upon users implementing the merging engine for specific languages. 
Unfortunately, its C/C++ parsers are either underdeveloped (i.e., accommodating only basic syntax) or unimplemented, rendering it unsuitable for real-world projects.
In addition, as mentioned in \S~\ref{subsec:limitation}, the machine learning approaches~\cite{zhang2022using, MergeBERT, elias2023towards, dongmerge, aldndni2023automatic, deepmerge, mergegen} meet the sequence length limitation challenge when facing the large-scale targets we chose.

\textit{Evaluation Environment.}
We evaluated \sys on an AMD EPYC 7713P server with 64/128 cores/threads, NVIDIA GeForce RTX 4090 (24GB GPU memory), and 256GB memory, running Ubuntu 20.04 with kernel 5.4.0.
%

\subsection{Effectiveness on Conflict Resolution}
\label{subs:effectiveness}
To address \textbf{RQ1}, we conducted a comparative analysis of the time consumption associated with two distinct approaches (Git merge w/ and w/o \sys) for merging conflict pairs.
As \sys only provides suggestions instead of resolutions of conflicts, We have to assign two types of technical staff to the manual merging phase:
\textbf{skilled staffs,} researchers with deep understanding of the project, and \textbf{ordinary staffs,} researchers having general understanding of the project.
Initially, the ordinary staff employed \sys to analyze the two versions and performed the merge utilizing the suggestions provided by \sys. 
Subsequently, the skilled staff manually merged the two candidate versions using Git merge without any additional assistance. 
We meticulously recorded the number of conflicts that can be analyzed by \sys and the time taken (with precision to the minute) from the commencement of conflict analysis to their complete resolution.

\begin{table}\small
\centering
\caption{\textmd{Conflict coverage of \sys. \textbf{Total Num} indicates the total number of conflicts in each dataset. \textbf{Analyzed Num} indicates the number of conflicts successfully analyzed by \sys. \textbf{Cover Rate} indicates the ratio of the above two columns.}}
\label{tb:conflict_effectivenss}
\begin{tblr}{
  width = \linewidth,
  colspec = {Q[223]Q[277]Q[202]Q[210]},
  hline{1,7} = {-}{0.08em},
  hline{2} = {-}{0.05em},
}
\textbf{Dataset} & \textbf{Analyzed Num} & \textbf{Total Num} & \textbf{Cover Rate} \\
Firefox          & 159                   & 160                & 99.38\%             \\
Linux Kernel     & 75                    & 88                 & 85.23\%             \\
LLVM             & 71                    & 72                 & 96.61\%             \\
PHP              & 236                   & 255                & 92.55\%             \\
MySQL            & 797                   & 964                & 82.68\%             
\end{tblr}            
\end{table}

Before comparing the effectiveness of merging w/ and w/o \sys, we first evaluate its ability to assess conflicts.
As shown in \autoref{tb:conflict_effectivenss}, we calculated the total number of conflicts for each dataset and recorded how many conflicts can be successfully analyzed by \sys.
\sys can give suggestions of more than 80\% conflicts from the five targets.
Notice that some conflicts cannot be handled by \sys, we manually dig the reasons behind the failure.
In the design section, we mentioned that \sys's analysis is based on LLVM IR.
However, in textual merging results, some conflicts are caused by the content of comments or preprocessing instructions (e.g., \textit{\#include} and macro definition).
Since comments do not belong to the code scope, and the preprocessing instructions will be expanded in the Clang front end, LLVM IR will lose the text information of these conflicts, and finally \sys will not give suggestions related to them.

\begin{figure}[h]
	\centerline{\includegraphics[width=\linewidth]{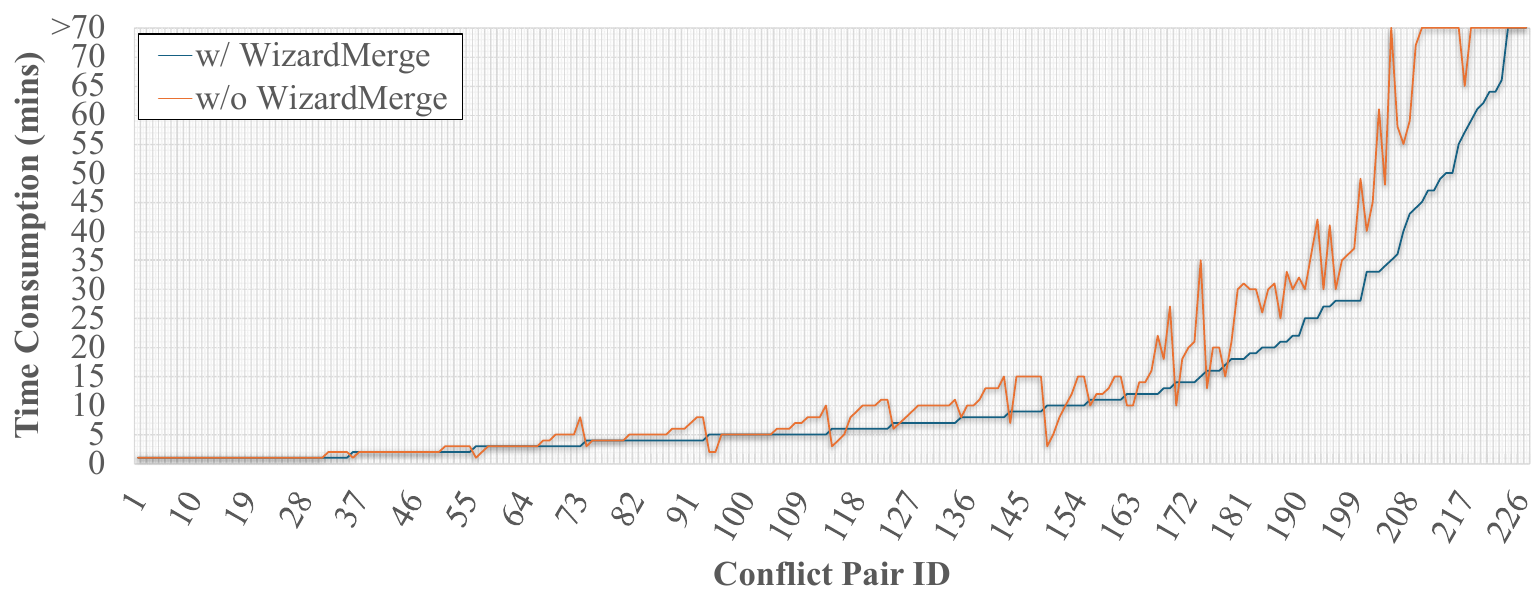}}
	\caption{\textmd{Manual merging time w/ and w/o \sys.}}
	\label{fg:time_consumption}
        \vspace{-\baselineskip}
\end{figure}

The effectiveness of merging w/ and w/o \sys is quantified by time consumption and is shown in \autoref{fg:time_consumption}.
When the time consumption of merging a pair of commits exceeds 70 minutes, we will mark it as "$>70$" and stop resolving it.
In 135 out of 277 conflict scenarios, \sys contributes to reducing version conflict resolution time, and as the time consumption w/o \sys increases, the assistance ability of \sys also becomes more obvious.
Remarkably, for the 205th conflict pair, manual merging w/o \sys costs more than 70 minutes, while with the help of \sys, the merging process can be finished in 35 minutes.
This exceptional performance is attributed to the Priority-Oriented Classification module in \sys, which provides categorization and fixes priority recommendations for manually resolved code blocks.
Simultaneously, the Violated DCB Detection module in \sys automatically identifies code blocks related to conflicts. 
This feature simplifies developers' efforts in analyzing dependencies and avoids the substantial time spent on the manual in-depth exploration of extensive code space required by Git merge.
However, \sys may not optimize the time consumption for all data points. 
In 18 out of 277 conflict cases, using \sys could even increase the burden of conflict resolution.
This occurs when only a small number of conflicts are present and few code blocks are affected, rendering \sys's assistance not substantial.
Additionally, as mentioned above, \sys fails to generate suggestions on comments or preprocessing instructions.

\begin{lstlisting}[label={lst:casestudy3}, language=diff, caption={\textmd{Case study for conflict \& violated DCB classification and \sys-produced resolving order assignment.}}]
diff --git a/gfx/layers/ScrollbarData.h b/gfx/layers/ScrollbarData.h
--- b/gfx/layers/ScrollbarData.h
+++ a/gfx/layers/ScrollbarData.h
@@ +38,10 -38,9 @@ struct ScrollbarData 
     * This constructor is for Thumb layer type.
     */
    ScrollbarData(ScrollDirection aDirection, float aThumbRatio,
<<<<<<< HEAD
+      OuterCSSCoord aThumbStart, OuterCSSCoord aThumbLength,
+      OuterCSSCoord aThumbMinLength, bool aThumbIsAsyncDraggable,
+      OuterCSSCoord aScrollTrackStart,
+      OuterCSSCoord aScrollTrackLength, uint64_t aTargetViewId)
=======
-      CSSCoord aThumbStart, CSSCoord aThumbLength,
-      bool aThumbIsAsyncDraggable, CSSCoord aScrollTrackStart,
-      CSSCoord aScrollTrackLength, uint64_t aTargetViewId)
>>>>>>> 281943a7a7564da09bd6076cac0c47a7a62144dc
        : mDirection(Some(aDirection)),
          mScrollbarLayerType(ScrollbarLayerType::Thumb),
          mThumbRatio(aThumbRatio),
@@ +65,16 -63,16 @@
   public:
    ScrollbarData() = default;
  
<<<<<<< HEAD
+ static ScrollbarData CreateForThumb(
+     ScrollDirection aDirection, float aThumbRatio, 
+     OuterCSSCoord aThumbStart,
+     OuterCSSCoord aThumbLength, OuterCSSCoord aThumbMinLength,
+     bool aThumbIsAsyncDraggable, 
+     OuterCSSCoord aScrollTrackStart,
+     OuterCSSCoord aScrollTrackLength, uint64_t aTargetViewId) {
=======
- static ScrollbarData CreateForThumb(ScrollDirection aDirection,
-     float aThumbRatio, CSSCoord aThumbStart,
-     CSSCoord aThumbLength,
-     bool aThumbIsAsyncDraggable,
-     CSSCoord aScrollTrackStart,
-     CSSCoord aScrollTrackLength,
-     uint64_t aTargetViewId) {
>>>>>>> 281943a7a7564da09bd6076cac0c47a7a62144dc
      return ScrollbarData(aDirection, aThumbRatio, aThumbStart, aThumbLength,
+        aThumbMinLength, aThumbIsAsyncDraggable,
+        aScrollTrackStart, aScrollTrackLength, aTargetViewId);
-        aThumbIsAsyncDraggable, aScrollTrackStart,
-        aScrollTrackLength, aTargetViewId);
    }

@@ MergeGuardian Result
@@ Group 0:
@@    |
@@    C-- ScrollbarData xxx/ScrollbarData.h 41-44, 41-43
@@      |
@@      C-- CreateForThumb xxx/ScrollbarData.h 68-74, 66-72
@@        |
@@        A-- CreateForThumb xxx/ScrollbarData.h 76-77
@@        |
@@        B-- (applied) CreateForThumb xxx/ScrollbarData.h 74-75
\end{lstlisting}

\autoref{lst:casestudy3} presents a case study illustrating how \sys classifies the conflicts and violated DCBs and assigns resolving privileges between \textbf{VA} \href{https://github.com/mozilla/gecko-dev/tree/145d3f680f74e0b7e22be9de4943baf4ff0c859b}{145d3f6} and \textbf{VB} \href{https://github.com/mozilla/gecko-dev/tree/281943a7a7564da09bd6076cac0c47a7a62144dc}{281943a}. 
Initially, both versions introduce distinct modifications to the \textit{ScrollbarData} constructor, resulting in a merge conflict (lines 8-17). 
Subsequently, both versions make different alterations to the parameter types of the \textit{CreateForThumb} method in the \textit{ScrollbarData} class, leading to another merge conflict (lines 25-41). 
However, Git's strategy reports these conflicts directly to the developer without providing additional clues.
In contrast, \sys utilizes dependency analysis between definitions to discern that the constructor of ScrollbarData will be called by \textit{CreateForThumb} (line 42). Consequently, \sys classifies the two conflicts into one group, assigning higher fix priority to lines 8-17 than to lines 25-41.
%

Additionally, VA and VB present diverse implementations of the function body of the \textit{CreateForThumb} method.
According to the three-way merging algorithm, Git applied the code snippets from VB (lines 45-46) after detecting these two conflicts.
Moreover, within the pair of DCBs represented by lines 43-46, although lines 45-46 are applied without conflicts, it cannot ensure the constructor of \textit{ScrollbarData} receives the correct parameter types when both conflicts are present. 
This is because, if the developer chooses to apply all VA modifications when resolving these conflicts, the applied lines 45-46 from VB will lack essential parameters, \textit{aThumbMinLength} and \textit{aThumbIsAsyncDraggabl}, used by \textit{ScrollbarData} constructor and the types of \textit{aThumbStart}, \textit{aThumbLength}, \textit{aScrollTrackStart}, and \textit{aScrollTrackLength} will not match the \textit{ScrollbarData} constructor prototype.
Nevertheless, \sys detects lines 45-46 and its mirror DCB (i.e., lines 43-44) through its innovative dependency analysis approach, treating them as violated DCBs. 
It reports them to developers along with other conflicts.
Since lines 43-46 depend not only on lines 8-17 (function call) but also on lines 25-41 (definition of prototype), these two violated DCBs are grouped with the previous two conflicts, with fixed priority coming after the two conflicts.
The resolution suggestions provided by \sys are shown in lines 49-58 of \autoref{lst:casestudy3}, where items prefixed with C indicate pairs of DCBs with conflicts, and items prefixed with A or B indicate DCBs from VA or VB considered as violated DCBs.


\begin{mdframed}[style=ccl]
\ccl
Our evaluation confirms \sys can give suggestions for 72.45\% of the conflicts.
We analyze the time expenditures associated with conflict resolution and their corresponding DCBs w/ and w/o the utilization of \sys to answering \textbf{RQ1}. 
On average, \sys demonstrates a 23.85\% reduction in resolution time. 
This underscores the efficiency and practicality of \sys in facilitating conflict resolution, even in the context of large-scale real-world software.
\end{mdframed}


\begin{table*}
\centering
\caption{\textmd{Violated DCB detection records. \textbf{CR VA} represents the LoC of CR violated DCB in VA, and \textbf{CR VB} represents the LoC of CR violated DCB in VB. The format $n/m$ indicates that there are $m$ LoCs in total, of which \sys detects $n$. Similarly, \textbf{CU VA} represents the LoC of CU violated DCB in VA, and \textbf{CU VB} represents the LoC of CU violated DCB in VB. The format $n/m$ indicates that \sys detect $m$ LoCs as CU violated DCB, of which $n$ are confirmed. \textbf{Recall} and \textbf{Precision} with $\dagger$ and $\ddagger$ stand for VA and VB respectively.}}
\label{tb:vDCB_detection}
\begin{tblr}{
  width = \linewidth,
  colspec = {Q[119]Q[96]Q[85]Q[88]Q[88]Q[104]Q[104]Q[121]Q[121]},
  hline{1,7} = {-}{0.08em},
  hline{2} = {-}{0.05em},
}
\textbf{Dataset}  & \textbf{CR-VA} & \textbf{CR-VB} & \textbf{CU-VA}   & \textbf{CU-VB}   & \textbf{Recall$\dagger$} & \textbf{Recall$\ddagger$} & \textbf{Precision$\dagger$} & \textbf{Precision$\ddagger$} \\
Firefox     & 151/174            & 112/123            & 75/652               & 98/717               & 88.30\%               & 91.06\%               & 11.50\%                  & 13.67\%                  \\
Linux Kernel & 107/131            & 156/192            & 171/1245             & 105/905              & 81.68\%               & 81.25\%               & 13.73\%                  & 11.60\%                  \\
LLVM         & 212/316            & 16/23              & \textit{\uline{N/A}} & \textit{\uline{N/A}} & 67.09\%               & 69.57\%               & \textit{\uline{N/A}}     & \textit{\uline{N/A}}     \\
PHP          & 92/132             & 49/139             & 226/324              & 213/358              & 69.70\%               & 35.25\%               & 69.75\%                  & 59.50\%                  \\
MySQL        & 12274/13100        & 3171/3469          & 1750/4597            & 1307/2427            & 93.69\%               & 91.41\%               & 38.07\%                  & 53.85\%                  
\end{tblr}
\vspace{-\baselineskip}
\end{table*}

\subsection{Violated DCB Detection}
\label{subsec:violatedDCBdetect}
To address both \textbf{RQ2} and \textbf{RQ3}, our initial step involves establishing the ground truth regarding the number of violated DCBs by manually merging two code versions using Git merge. 
Subsequently, we employ \sys to conduct an analysis and document the identified violated DCBs. 
As not all code segments within a DCB pertain to violated DCBs, we utilize \textbf{Line of Code (LoC)} for careful quantification of the violated DCBs. 
These violated DCBs are further categorized into \textbf{Conflict-Related (CR)} and \textbf{Conflict-Unrelated (CU)} violated DCBs.

\subsubsection{Conflict-Related Violated DCBs}
In addition to addressing conflict DCBs, developers should reassess the DCBs automatically applied by Git merge. 
Resolving one conflict might potentially impact other sets of DCBs. 
Throughout our evaluation, we recorded the LoC for DCBs linked to conflicts in both versions. 
This data reflects \sys's capability in elucidating the correlation between conflicts and violated DCBs.
\autoref{tb:vDCB_detection} illustrates the CR-violated DCB detection results of \sys.
As VA and VB's codes are different, the LoC of CR-violated DCBs in the same diff pair may vary as well.
Thus, we documented the LoC from both VA and VB.
According to the evaluation, 167 out of 227 conflict scenarios are confirmed to contain CR-violated DCBs.
For most of the targets, \sys successfully detected more than 60\% CR-violated DCBs, providing a promising assistive cue for developers in executing manual fixes and facilitating them in exploring the CR-violated DCBs.
Noticeably, although MySQL datasets have the most LoC of CR-violated DCBs, \sys's detection recall even reaches 93.69\%.
For LLVM and PHP, \sys showed moderate recalls.

We also delved into the reasons behind \sys's inability to detect CR-violated DCBs in certain datasets. 
Firstly, the Linux Kernel encountered compilation failure with the "O0" option, a crucial setting for preserving debug information and preventing code optimization. 
Although "Og" can be used for better debug ability, it is akin to "O1," resulting in the elimination of significant debug information~\cite{linuxkernel-o0}. 
For instance, inlined functions are inserted at the invocation location and subsequently removed from the original code. 
As \sys depends on the collaboration of debug information and LLVM IR, the absence of debug information leads to the failure of node generation and DCB matching. 
Secondly, some datasets include violated dependencies among local variables within a specific function or macro definition, which \sys's analysis currently does not support.

\begin{lstlisting}[label={lst:casestudy1}, language=diff, caption={\textmd{Case study for conflict-related violated DCBs}}]
diff --git a/drivers/net/virtio_net.c b/drivers/net/virtio_net.c
--- b/drivers/net/virtio_net.c
+++ a/drivers/net/virtio_net.c
@@ -126,12 +127,9 @@ struct virtnet_stat_desc virtnet_rq_stats_desc[] = {
 #define VIRTNET_SQ_STATS_LEN	ARRAY_SIZE(virtnet_sq_stats_desc)
 #define VIRTNET_RQ_STATS_LEN	ARRAY_SIZE(virtnet_rq_stats_desc)
 
<<<<<<< 5c5e0e81202667f9c052edb99699818363b19129
- struct virtnet_rq_dma {
-   dma_addr_t addr;
-   u32 ref;
-   u16 len;
-   u16 need_sync;
=======
+ struct virtnet_interrupt_coalesce {
+   u32 max_packets;
+   u32 max_usecs;
>>>>>>> 1acfe2c1225899eab5ab724c91b7e1eb2881b9ab
 };
@@ -147,6 +145,8 @@ struct send_queue {
 
 	struct virtnet_sq_stats stats;
 
+   struct virtnet_interrupt_coalesce intr_coal;
 
 	struct napi_struct napi;
@@ -164,6 +164,8 @@ struct receive_queue {
 
 	struct virtnet_rq_stats stats;
 
+   struct virtnet_interrupt_coalesce intr_coal;

 	/* Chain pages by the private ptr. */
@@ -185,12 +187,6 @@ struct receive_queue {
 	struct xdp_rxq_info xdp_rxq;

-   /* Record the last dma info to free ... */
-   struct virtnet_rq_dma *last_dma;
-   /* Do dma by self */
-   bool do_dma;
 };
@@ -295,10 +292,8 @@ struct virtnet_info {
 	u32 speed;
 
 	/* Interrupt coalescing settings */
-   u32 tx_usecs;
-   u32 rx_usecs;
-   u32 tx_max_packets;
-   u32 rx_max_packets;
+   struct virtnet_interrupt_coalesce intr_coal_tx;
+   struct virtnet_interrupt_coalesce intr_coal_rx;

\end{lstlisting}

%
\autoref{lst:casestudy1} presents an intriguing case study revealing conflict-related violated DCBs detected by \sys between \textbf{VA} \href{https://github.com/git/git/tree/5c5e0e81202667f9c052edb99699818363b19129}{5c5e0e8} and \textbf{VB} \href{https://github.com/git/git/tree/1acfe2c1225899eab5ab724c91b7e1eb2881b9ab}{1acfe2c}. 
In VA, the definition for \textit{virtnet\_rq\_dma} is present, whereas at the same position in VB, \textit{virtnet\_interrupt\_coalesce} is defined.
Git identifies these versions' different modifications as a conflict and prompts manual resolution for developers. 
Subsequently, Git designates other DCBs as not in conflict and applies them based on their consistency with the base version's DCBs. 
Eventually, the code snippets from lines 24, 31, and 50-51 (from VA) along with 37-40 (from VB) are applied.
Despite the differences between VA and VB in lines 46-51, lines 46-49 from VB are consistent with the base version's snippets. 
According to the three-way merging strategy, Git prefers to apply lines 50-51 from VA.
In contrast, \sys analyzes all modified code and discovers that even though the conflict is highlighted, some related DCBs have been applied without alerting the developers. 
For instance, applying lines 37-40 from VB assumes the presence of struct \textit{virtnet\_rq\_dma}. 
Yet, if the developer opts for lines 15-17 as the conflict resolution, the structure becomes undefined within that context. 
Similarly, applying lines 24, 31, and 50-51 from VA assumes the definition of struct \textit{virtnet\_interrupt\_coalesce}, while they will cause compilation errors if the developer chooses lines 9-13 as the conflict resolution.
Consequently, \sys identifies these code snippets as violated DCBs in conjunction with the conflict. 
Furthermore, \sys assigns higher priority to the conflict, signifying the dependency of other DCBs on the conflict and suggesting developers resolve the conflict before reconsidering these violated DCBs.
%

\subsubsection{Conflict-Unrelated Violated DCBs}
Throughout this evaluation, we document the LoC for DCBs unrelated to conflicts in both variant versions reported by \sys. 
Following this, we manually inspected the identified DCBs and calculated the number of LoC genuinely violated among them.

The results of \sys's detection of CU-violated DCBs are also presented in \autoref{tb:vDCB_detection}. 
As CU-violated DCBs are isolated from conflicts, \sys demonstrates the capability to detect them, even in scenarios where it fails to handle any conflicts. 
In total, \sys detected CU-violated DCBs in 137 out of 227 conflict scenarios, while no CU-violated DCBs were found in LLVM.
These DCBs, lacking necessary dependent definitions and automatically applied by Git merge without developer notification, do not lead to conflicts but can result in compile errors or even latent bugs affecting users at runtime.
While \sys showcases its potential in uncovering violated dependency issues, the evaluation also highlights instances of \textbf{false positive} CU-violated DCBs. 
This arises from the lack of function body analysis in \sys. 
According to \sys's DCB detection algorithm \S \ref{ssc:vioaltedDCBDetection}, the relationship between two DCBs from the same function relies on line numbers, with the latter one forcibly dependent on the former one. 
If the two DCBs are applied from different versions, \sys marks them as CU-violated DCBs, even if they are unrelated. 
This conservative algorithm may result in more false positives, but \sys prioritizes its application to prevent overlooking the detection of CU-violated DCBs.

 

 
 

\begin{figure*}[t]
\begin{minipage}[t]{0.45\textwidth}
\begin{lstlisting}[label={lst:casestudy2A}, language=C++, caption={\textmd{The definitions of function \textit{EnsureNSSInitializedChromeOrContent} are the same in both versions}}]
// security/manager/ssl/nsNSSComponent.cpp
bool EnsureNSSInitializedChromeOrContent() {
    ...
    if (XRE_IsParentProcess()) {
    // Create nsISupports * via template class nsCOMPtr. 
        nsCOMPtr<nsISupports> nss = 
            do_GetService(PSM_COMPONENT_CONTRACTID);
        if (!nss) {
          return false;
        }
        initialized = true;
        return true;
    }
  ...
}
\end{lstlisting}
\end{minipage}
\hspace{0.05\linewidth}
\begin{minipage}[t]{0.45\textwidth}
\begin{lstlisting}[label={lst:casestudy2B}, language=diff, caption={\textmd{Difference between commits d2956560d539 and a20620423a53 in file xpcom/base/nsCOMPtr.h}}]
diff --git a/xpcom/base/nsCOMPtr.h b/xpcom/base/nsCOMPtr.h
--- a/xpcom/base/nsCOMPtr.h
+++ b/xpcom/base/nsCOMPtr.h
template <class T>
class MOZ_IS_REFPTR nsCOMPtr final {
    ...
};

...

+ template <>
+ class MOZ_IS_REFPTR nsCOMPtr<nsISupports> 
+                       : private nsCOMPtr_base {
+   ...
+ }

\end{lstlisting}
\end{minipage}
\vspace{-1.3\baselineskip}
\end{figure*}

%
\autoref{lst:casestudy2A} and \autoref{lst:casestudy2B} present another case study where \sys explores conflict-unrelated violated DCBs within Firefox (between versions \textbf{VA} \href{https://github.com/mozilla/gecko-dev/tree/d2956560d539c54eaf56297a27b139b862ac858f}{d295656} and \textbf{VB} \href{https://github.com/mozilla/gecko-dev/tree/a20620423a5363cf7afdac81e062bc687c29366a}{a206204}), which were overlooked by Git merge. 
Specifically, in the file \textit{nsNSSComponent.cpp}, both versions define the function \textit{EnsureNSSInitializedChromeOrContent} with the same contents.
In \autoref{lst:casestudy2A} lines 6-7, a pointer to \textit{nsISupports} is created via the template class \textit{nsCOMPtr}. 
The definition of \textit{nsCOMPtr} is found in \textit{nsCOMPtr.h}. However, in addition to the generic template, a template specialization~\cite{templatec++} is also defined for \textit{nsISupports} in VB, as shown in lines 11-15 in \autoref{lst:casestudy2B}. 
This specialization of \textit{nsCOMPtr} for \textit{nsISupports} allows users to employ \textit{nsCOMPtr<nsISupports>} similarly to how they use \textit{nsISupports*} and \textit{void*}, essentially as a 'catch-all' pointer pointing to any valid [XP]COM interface \cite{xpcom}.
When attempting to merge \textit{nsCOMPtr.h} from the two versions, Git applies the file from VA because the file from VB is identical to the file from the base version. 
Consequently, in the function \textit{EnsureNSSInitializedChromeOrContent}, the usage of \textit{nsCOMPtr<nsISupports>} will follow lines 4-7 in \autoref{lst:casestudy2B}, causing \textit{nss} to only be able to point to the single \textit{[XP]COM-correct nsISupports} instance within an object. 
This inconsistency can lead to a potential \textbf{type confusion} bug \cite{typescan, hextype}.
Since no compilation error occurs after successfully being applied by Git merge, it would be extremely challenging for developers to notice such a bug, not to mention identify the root cause.
With the assistance of \sys, the loss of template specialization can be detected in the merged version, thereby notifying developers to address it manually.

%

%
\begin{mdframed}[style=ccl]
\ccl
We assess the efficacy of \sys in identifying CR-violated and CU-violated DCBs by quantifying these DCBs in terms of Lines of Code (LoC), addressing both \textbf{RQ2} and \textbf{RQ3}. 
The results reveal that \sys achieves a detection recall of 80.09\% and 73.71\% for CR-violated DCBs from both merging candidates in the five target projects on average.
Furthermore, despite attaining precision rates of 33.26\% and 34.66\% for CU-violated DCBs in the four target projects on average, \sys effectively reveals deeply concealed CU-violated DCBs that are overlooked by developers. 
To underscore the impacts of CR-violated and CU-violated DCBs, we present case studies that further elucidate the rationale behind \sys's noteworthy findings.
\end{mdframed}

\section{Discussion}

In this section, we discuss the present limitations in \sys's design and explore potential avenues for enhancing \sys.

\textit{Improve Graph Generation.}
As discussed in \S \ref{subsec:rsg}, \sys incorporates three node types and five edge types to construct the ODG. 
While it adeptly manages numerous merging scenarios, the evaluation outcomes highlighted in \S \ref{sec:eval} suggest that this incompleteness could result in an inability to address certain conflicts or violated DCB instances.
For example, \sys cannot infer the dependency from a global variable to a function if the global variable is initialized by the function.
This is because, during the compilation process, the initial values of global variables are established before the execution. 
In contrast, a function's return value is accessible only after being executed. 
When initializing global variables with a function's return value, the compiler transforms the function invocation into initialization code.
To ensure the execution of this initialization code before the main program commences, the compiler positions this code within a specific section of the executable file, commonly denoted as the ".init" segment. 
This separation facilitates the execution of these initialization routines prior to the initiation of the main program logic.
We leave updating the features as our future work to adapt to such scenarios.

\textit{Assess Code Optimization.}
\sys conducts dependency analysis based on the compilation metadata.
Nevertheless, it overlooks a substantial amount of code information due to compilation optimizations like unused function elimination and vectorization~\cite{llvm_opt}. 
As an interim measure, \sys presently enforces the "O0" option. 
However, some projects require compilation optimization for successful building. 
For instance, the Linux kernel relies on optimization to eliminate redundant code segments and disallows the "O0" option.
In the future, we aim to explore how to adapt our approach to arbitrary optimization options.
\section{Related Work}

%

\subsection{Merging Strategies}

\paragraph{Structured Merging}
Mastery~\cite{mastery} introduces an innovative structured merging algorithm that employs a dual approach: top-down and bottom-up traversal. 
It can minimize redundant computations and enable the elegant and efficient handling of the code that has been relocated from its original position to another location.
Spork~\cite{spork} extends the merging algorithm from the 3DM ~\cite{lindholm2004three} to maintain the well-structured code format and language-specific constructs.
To fully retain the original functionality of the merged code, SafeMerge~\cite{safemerge} leverages a verification algorithm for analyzing distinct edits independently and combining them to achieve a comprehensive demonstration of conflict freedom.

\paragraph{Semi-structured Merging}
JDime~\cite{jdime} introduces a structured merge approach with auto-tuning that dynamically adjusts the merge process by switching between unstructured and structured merging based on the presence of conflicts.
FSTMerge~\cite{fstmerge} stands as an early example of semi-structured merging, offering a framework for the development of semi-structured merging.
Intellimerge~\cite{intellimerge} transforms source code into a program element graph via semantic analysis, matches vertices with similar semantics, merges matched vertices' textual content, and ultimately translates the resulting graph back into source code files.

The above works, however, cannot help the developers with resolving conflicts.
On the contrary, \sys is constructed atop Git merge, furnishing developers with clues to aid manual resolution instead of dictating final merging decisions.
\vspace{-0.5\baselineskip}

\subsection{Conflict Resolution}


\paragraph{Conflict Resolution Assistance}
TIPMerge~\cite{tipmerge} evaluates the developers' experience with the key files based on the project and branch history~\cite{tipmerge} to determine the most suitable developers for merging tasks. 
Automerge~\cite{automerge} relies on version space algebra~\cite{lau2003programming} to represent the program space and implement a ranking mechanism to identify a resolution within the program space that is highly likely to align with the developer's expectations. 
SoManyConflicts~\cite{somanyconflicts} is implemented by modeling conflicts and their interrelations as a graph and employing classical graph algorithms to offer recommendations for resolving multiple conflicts more efficiently. 

\paragraph{Machine Learning Approaches}
MergeBERT~\cite{MergeBERT} reframes conflict resolution as a classification problem and leverages token-level three-way differencing and a transformer encoder model to build a neural program merge framework. 
Gmerge~\cite{zhang2022using} investigates the viability of automated merge conflict resolution through k-shot learning utilizing pre-trained large language models (LLMs), such as GPT-3~\cite{floridi2020gpt}. 
MergeGen~\cite{mergegen} treats conflict resolution as a generation task, which can produce new codes that do not exist in input, and produce more flexible combinations.

The essential difference compared with existing assistance systems and machine learning approaches is that \sys considers all modified code snippets, which further helps developers resolve the violated DCBs in merging scenarios.

\section{Conclusion}

We introduce \sys, a novel code-merging assistance system leveraging the merging results of Git and dependency analysis based on LLVM IR, named definition range, and debug information.
Via the dependency analysis, \sys is able to detect the loss of dependency of the named definitions, therefore providing more accurate merging order suggestions including conflicts and applied code blocks for the developers.
Our evaluation demonstrates that \sys significantly diminishes conflict merging time costs.
Beyond addressing conflicts, \sys provides merging suggestions for most of the code blocks potentially affected by the conflicts. 
Moreover, \sys exhibits the capability to identify conflict-unrelated code blocks which should require manual intervention yet automatically applied by Git.

\clearpage


\bibliographystyle{ACM-Reference-Format}
\bibliography{cite}


\end{document}